\documentclass[aps,10pt,prl,showpacs,twocolumn, superscriptaddress]{revtex4-1}
\usepackage{graphicx,graphics}
\usepackage[export]{adjustbox}
\usepackage{amssymb}
\usepackage{amsfonts}
\usepackage{color}
\usepackage{bm}

\newcommand\varpm{\mathbin{\vcenter{\hbox{%
  \oalign{\hfil$\scriptstyle+$\hfil\cr
          \noalign{\kern-.3ex}
          $\scriptscriptstyle({-})$\cr}%
}}}}
\newcommand\varmp{\mathbin{\vcenter{\hbox{%
  \oalign{$\scriptstyle({+})$\cr
          \noalign{\kern-.3ex}
          \hfil$\scriptscriptstyle-$\hfil\cr}%
}}}}
\DeclareMathAlphabet      {\mathbf}{OT1}{cmr}{bx}{n}

\begin{document}
\title{Large Negative Quantum Renormalization of Excitation Energies in the Spin-1/2 Kagome Lattice Antiferromagnet Cs$_2$Cu$_3$SnF$_{12}$}
\author{T.~Ono}
\affiliation{Department of Physical Science, School of Science, Osaka Prefecture University, Sakai, Osaka 599-8531, Japan}
\author{K.~Matan}
\affiliation{Department~of~Physics,~Faculty~of~Science,~Mahidol~University, Bangkok 10400, Thailand}
\affiliation{ThEP,~Commission of Higher Education,~Bangkok 10400, Thailand}
\author{Y.~Nambu} 
\affiliation{IMRAM, Tohoku University, Sendai, Miyagi 980-8577, Japan}
\author{T.~J.~Sato}
\affiliation{IMRAM, Tohoku University, Sendai, Miyagi 980-8577, Japan}
\author{K. Katayama} 
\affiliation{Department of Physics, Tokyo Institute of Technology, Meguro-ku, Tokyo 152-8551, Japan}
\author{S. Hirata}
\affiliation{Department of Physics, Tokyo Institute of Technology, Meguro-ku, Tokyo 152-8551, Japan}
\author{H.~Tanaka}
\email{tanaka@lee.phys.titech.ac.jp}
\affiliation{Department of Physics, Tokyo Institute of Technology, Meguro-ku, Tokyo 152-8551, Japan}

\date{\today}
\begin{abstract}
Magnetic excitations in the spin-$\frac{1}{2}$ distorted kagome lattice antiferromagnet Cs$_2$Cu$_3$SnF$_{12}$, which has an ordered ground state owing to the strong Dzyaloshinskii-Moriya interaction, were studied using inelastic neutron scattering. Although the spin-wave dispersion can be qualitatively understood in terms of linear spin-wave theory (LSWT), the excitation energies are renormalized by a factor of approximately 0.6 from those calculated by LSWT, almost irrespective of the momentum transfer.  This inadequacy of LSWT, which is attributed to quantum fluctuations, provides evidence of negative quantum renormalization in the spin-$\frac{1}{2}$ kagome lattice antiferromagnet.
\end{abstract}
\maketitle

Ubiquitous magnetic excitations in conventional magnets with the N\'eel state are generally well described by LSWT. In low-dimensional quantum magnets, however, dominant quantum effects significantly modify the magnetic excitations. In particular, for an $S\,{=}\,1/2$ antiferromagnetic Heisenberg spin chain, the exact spinon excitation energies are larger than that calculated using LSWT by a factor of ${\pi}/2$\cite{dCP}, which was verified through an inelastic neutron scattering experiment on the spin-1/2 one-dimensional (1D) Heisenberg antiferromagnet CuCl$_2\,{\cdot}\,2$(C$_5$D$_5$)\cite{Endoh}. This quantum enhancement of excitation energies is known as the quantum renormalization.

The spin-1/2 2D kagome-lattice antiferromagnet (KLAF) is a research frontier with the potential to realize a disordered ground state arising from the synergistic effect of strong frustration and quantum fluctuations\cite{Sachdev,Waldtmann,Nakano,Wang,Hermele,Yan,Depenbrock,Nishinoto,Singh,Evenbly,Hwang1}. The theoretical consensus for the case of Heisenberg spins is that the classical N\'eel state, which is robust in conventional magnets, is supplanted by a disordered quantum state. However, the nature of the ground state, which is the basis for  the discussion of excitations, has not been theoretically elucidated. Innovative theoretical studies have been conducted on the spin-1/2 nearest-neighbor Heisenberg KLAF using a variety of approaches. Most of the recent results suggest the existence of nonmagnetic ground states described by spin liquids\cite{Wang,Hermele,Yan,Depenbrock,Nishinoto} and valence-bond solids\cite{Singh,Evenbly,Hwang1}.  Experimentally, the lack of an ideal model has hindered detailed studies of intrinsic excitations of kagome magnets. Nevertheless, great effort has been made to search for approximate realizations of the spin-1/2 KLAF, which exhibits a diversity of states\cite{Shores,Helton,Han,Mueller,Hiroi,Okamoto,Ono,Morita,Matan1}.

\begin{figure}[htb]
\centering \vspace{0in}
\includegraphics[width=5.5cm]{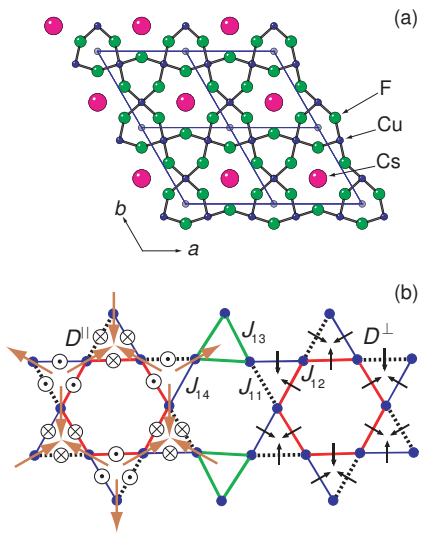}
\caption{(Color online) (a) Crystal structure at room temperature viewed along the $c$ axis, where fluorine ions located outside the kagome layer are omitted. Thin lines denote the unit cell. (b) Diagram showing the connectivity of $S\,{=}\,1/2$ Cu$^{2+}$ spins via the nearest-neighbor exchange interactions $J_{11}$, $J_{12}$, $J_{13}$ and $J_{14}$. Configurations of the out-of-plane component $D^{\|}$ and in-plane component $D^{\bot}$ of the DM vectors, deduced from the highly symmetric room-temperature structure, are illustrated on the left and right, respectively. The large arrows on the left indicate the ${\bm q}=0$ structure assumed in the LSWT calculations.}
\vspace{-3mm}
\label{fig1}
\end{figure}

\begin{figure}
\centering \vspace{0in}
\includegraphics[width=8.5cm]{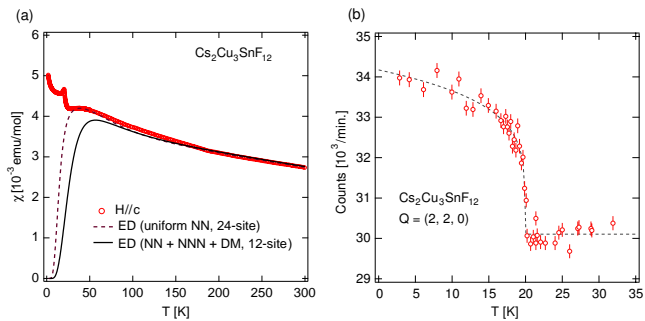}
\caption{(Color online) (a) Temperature dependence of magnetic susceptibility in Cs$_2$Cu$_3$SnF$_{12}$. Dashed line denotes the result obtained by exact diagonalization for a 24-site uniform kagome cluster with $J\,{=}\,20.7$ meV and $g\,{=}\,2.49$, while solid line is the result obtained for a 12-site distorted kagome cluster with $J_{\rm avg}\,{=}\,19.8$ meV and the same interaction coefficients $a_i$, $J_2/J_1$ and $d_z$ as those obtained from the analysis of spin-wave dispersions with $d_{\rm p}\,{=}\,0$ and $g\,{=}\,2.43$. (b) Temperature dependence of the magnetic Bragg reflection at ${\bm Q}\,{=}\,(2, 2, 0)$. The dashed line serves as a visual guide.}
\vspace{-3mm}
\label{fig2}
\end{figure}

A$_2$Cu$_3$SnF$_{12}$ (A=Rb, Cs) is a promising family of spin-1/2 KLAFs\cite{Ono,Morita}. Rb$_2$Cu$_3$SnF$_{12}$ has a distorted kagome lattice and a gapped $S\,{=}\,0$ singlet ground state\cite{Ono,Morita,Matan1}. A study of singlet-to-triplet excitations in Rb$_2$Cu$_3$SnF$_{12}$ using inelastic neutron scattering revealed a pinwheel motif of strongly interacting dimers\cite{Morita,Matan1,Matan2}. All relevant spin Hamiltonian parameters were determined, which suggested the dominant effect of the Dzyaloshinskii-Moriya (DM) interaction\cite{Morita,Matan1,Matan2,Hwang2}. On the other hand, Cs$_2$Cu$_3$SnF$_{12}$ has a uniform kagome lattice at room temperature with the lattice parameters $a\,{=}\,7.142(4)$~\AA~and $c\,{=}\,20.381(14)$~\AA~\cite{Ono}, as shown in Fig.~\ref{fig1}(a). This compound undergoes a structural transition at $T_{\rm s}\,{=}\,185$ K and magnetic ordering at $T_{\rm N}\,{=}\,20.0$ K~\cite{Ono}. The magnetic susceptibility exhibits a small anomaly at $T_{\rm s}$ and a large increase at $T_{\rm N}$ (Fig.~\ref{fig2}(a)). The presence of superlattice reflections below $T_{\rm s}$ suggests the doubling of the in-plane lattice parameter, giving rise to a $2a\,{\times}\,2a$ enlarged unit cell.  Above $T_{\rm N}$, the magnetic susceptibility is in good agreement with the theoretical susceptibility obtained from exact diagonalization for the 24-site kagome cluster\,\cite{Misguich} (Fig.~\ref{fig2}(a)). This suggests that the exchange network remains approximately uniform. 

Low-energy magnetic excitations in the spin-1/2 distorted KLAF Cs$_2$Cu$_3$SnF$_{12}$ can be described by the collective disturbance of the ordered moments. Although these magnetic excitations in the classical spin-5/2 KLAF KFe$_3$(OH)$_6$(SO$_4$)$_2$ are well described by LSWT\cite{Matan3,Yildirim}, little is known about the quantum effect for the spin-1/2 case, where large quantum renormalization is expected to emerge. In this letter, we present the first evidence of the large negative renormalization of spin-wave energies with respect to the LSWT result in Cs$_2$Cu$_3$SnF$_{12}$. This observation provides a striking contrast to the well-known positive quantum renormalization of excitation energies in the $S\,{=}\,1/2$ antiferromagnetic Heisenberg spin chain,\cite{dCP} for which the renormalization factor is exactly ${\pi}/{2}$. 

Cs$_2$Cu$_3$SnF$_{12}$ crystals were synthesized in accordance with the chemical reaction 2CsF + 3CuF$_2$ + SnF$_4 \rightarrow$ Cs$_2$Cu$_3$SnF$_{12}$. CsF, CuF$_2$ and, SnF$_4$ were dehydrated by heating in vacuum at about 100$^\circ$C. First the materials were packed into a Pt tube of 9.6 mm inner diameter and 100 mm length in the ratio of $3\,{:}\,3\,{:}\,2$. One end of the Pt tube was welded and the other end was tightly folded with pliers and placed between Nichrome plates. Single crystals were grown from the melt. The temperature of the furnace was lowered from 850 to 750$^\circ$C over 100 hours. After collecting the well-formed pieces of crystal, we repeated the same procedure.  Inelastic neutron scattering measurements were performed on two co-aligned single crystals of Cs$_2$Cu$_3$SnF$_{12}$ (total mass of 3.3 g) with a sample mosaic of about $1^\circ$ at GPTAS and HER, which are triple-axis spectrometers run by the Institute for Solid State Physics, University of Tokyo. At GPTAS, the final energy of the thermal neutrons was fixed at 14.7 meV. The collimations were $40'\,{-}\,40'\,{-}\,\textrm{sample}\,{-}\,40'\,{-}\,80'$. A pyrolytic graphite (PG) filter was placed after the sample to remove contamination from higher-order neutrons. The vertically focused (horizontally flat) PG crystals were used to analyze the scattered neutrons. At HER, the final energy of the cold neutrons was fixed at 5 meV. The scattered neutrons were analyzed using the central three blades of a seven-blade doubly focused PG analyzer. A cool Be or oriented-PG-crystal filter was placed in the incident beam and a room-temperature Be filter was placed in the scattered beam. In the analysis of the HER data, effective collimations of $10'\,{-}\,40'\,{-}\,\textrm{sample}\,{-}\,160'\,{-}\,120'$ were used. For both experiments, the sample was aligned with the $(h, k, 0)$ plane horizontal to measure spin-wave excitations within the kagome plane. The sample was cooled to the base temperature of 3 K using a $^4$He closed cycle cryostat.

Using the $2a\,{\times}\,2a$ enlarged unit cell for the low-temperature crystal structure, we observed an increased scattering intensity due to magnetic Bragg reflections below $T_{\rm N}\,{=}\,20.0$ K at ${\bm Q}_{\rm m}\,{=}\,(2m, 2n, 0)$, where $m$ and $n$ are integers. The ordering wave vectors correspond to the reciprocal lattice points of the uniform kagome lattice above $T_{\rm s}\,{=}\,185$ K. Figure~\ref{fig2}(b) shows the temperature dependence of the magnetic Bragg reflection at ${\bm Q}\,{=}\,(2, 2, 0)$. The scattering intensity above $T_{\rm N}\,{=}\,20.0$~K arises from a nuclear reflection. This result indicates that the ordered state has a ${\bm q}\,{=}\,0$ structure. Hence the center of the 2D Brillouin zone located at ${\bm Q}_{\rm m}$ is expected to give rise to strong spin-wave scattering.

\begin{figure*}[htb]
\centering \vspace{0in}
\includegraphics[width=15.0cm]{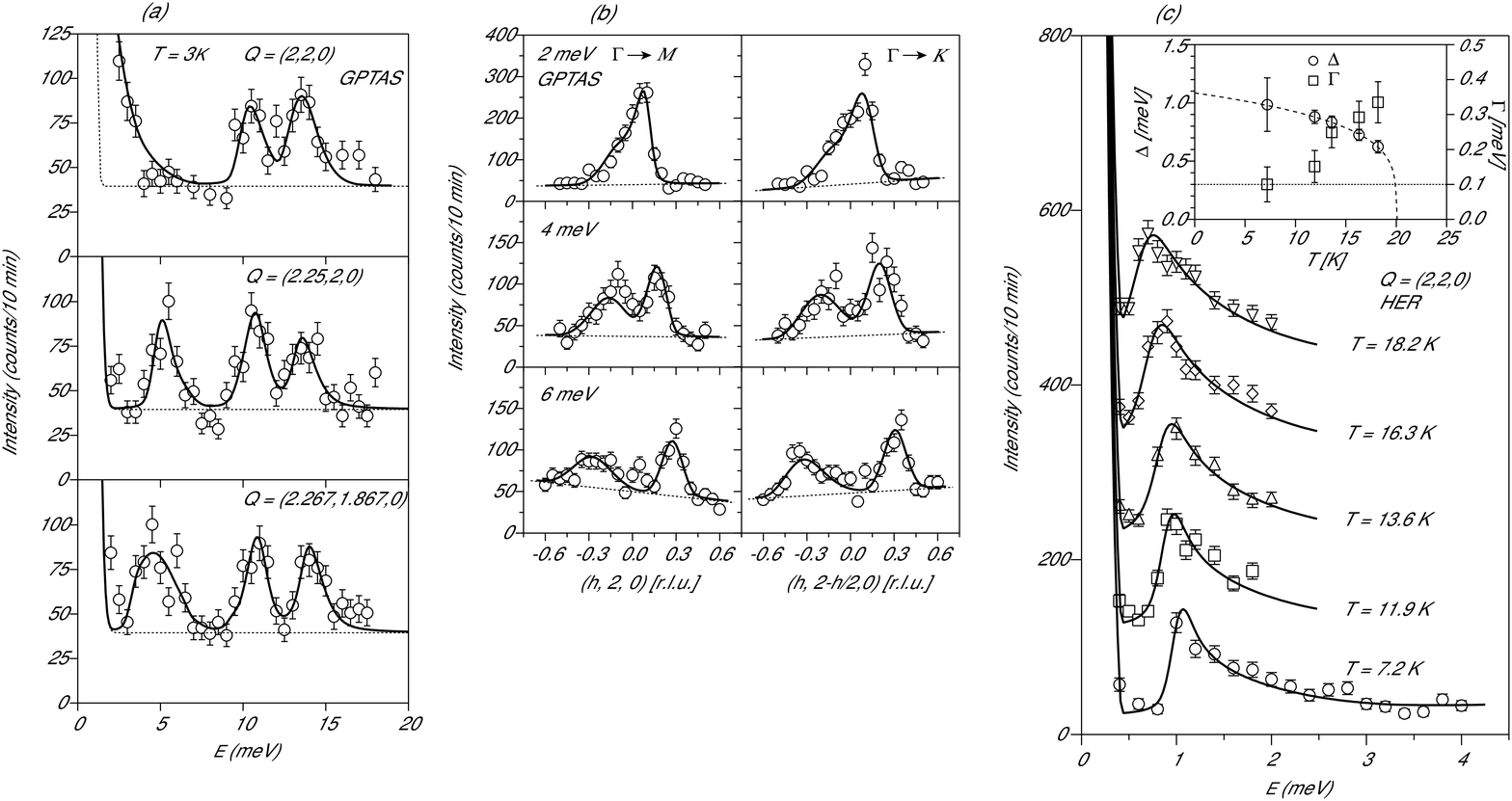}
\caption{(a) Constant-${\bm Q}$ scans measured at ${\bm Q}\,{=}\,(2, 2, 0)$, $(2.25, 2, 0)$, and $(2.267, 1.867, 0)$. (b) Constant-energy scans measured at ${\hbar}{\omega}\,{=}\,2$, $4$, and $6$ meV along two independent high-symmetry directions (see Fig.~\ref{fig4}(c)). (c) Temperature dependence of the spin gap at the $\Gamma$-point. The main panel shows constant-${\bm Q}$ scans measured at the $\Gamma$-point at different temperatures. Data sets for different temperature are shifted vertically by 100. The inset shows the temperature dependences of the spin-gap energy $\Delta$ and peak width $\Gamma$. The dotted line denotes the resolution of the instrument obtained by the convolution fitting, and the dashed lines serve as a visual guide. The error bar denotes the statistical error.}
\vspace{-3mm}
\label{fig3}
\end{figure*}

\begin{figure}[tbh]
\centering \vspace{0in}
\includegraphics[width=7.0cm]{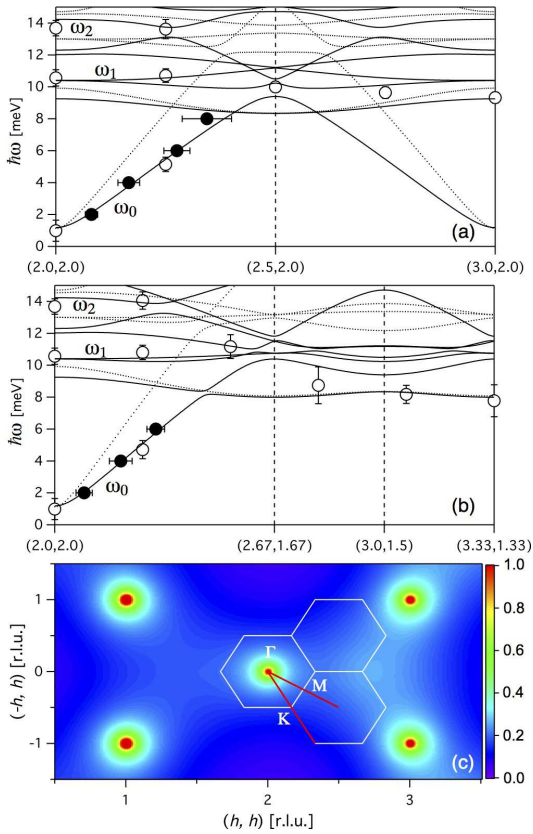}
\caption{(Color online) (a) and (b) Experimental data and calculated spin-wave dispersions along the two high-symmetry directions denoted by thick red lines in (c). Open symbols indicate the data measured around $(2, 2, 0)$ while closed symbols indicate the data measured at the equivalent point around $(0, 2, 0)$. Solid lines denote the best fit obtained using the DM model in eq. (\ref{eq1}), and dotted lines denote dispersions with $J_{\tiny\textrm{avg}}^{\tiny\textrm{mag}}\,{=}\,19.8$ meV obtained from the magnetic susceptibility, $J_2\,{=}\,-1.07$ meV, $d_z\,{=}\,-0.18$, and $d_{\rm p}\,{=}\,0.033$. (c) Calculated energy-integrated scattering intensity of Cs$_2$Cu$_2$SnF$_{12}$.}
\vspace{-3mm}
\label{fig4}
\end{figure}

Figure~\ref{fig3}(a) shows constant-${\bm Q}$ scans measured using the GPTAS spectrometer. The scans were performed at 3 K and at three different momentum transfers ${\bm Q}\,{=}\,(2, 2, 0)$, $(2.25, 2, 0)$, and $(2.267, 1.867, 0)$. At the zone center ($\Gamma$-point) ${\bm Q}\,{=}\,(2, 2, 0)$, we clearly observed two spin-wave excitations at 10.7(5) meV and 13.6(4) meV, and extra scattering above the background below 5 meV (top panel of Fig.~\ref{fig3}(a)). A high-resolution measurement using the cold-neutron spectrometer HER revealed a spin gap of 1.0(6) meV as shown in Fig.~\ref{fig3}(c). Away from the zone center along the $\Gamma\,{\rightarrow}\,{\rm M}$ and $\Gamma\,{\rightarrow}\,{\rm K}$ directions, we clearly observed three peaks representing three branches of spin-wave excitations, as shown in the middle and bottom panels of Fig.~\ref{fig3}(a), respectively. Figure~\ref{fig3}(b) shows constant-energy scans taken along two independent high-symmetry directions from the $\Gamma$-point to the M- and K-points (Fig.~\ref{fig4}(c)). For both constant-${\bm Q}$ and constant-energy scans, the peak width is resolution-limited and the line shape is well described by the convolution with the resolution function. As the temperature increases toward $T_{\rm N}$, the energy of the spin gap $\Delta$, which scales with the order parameter, decreases toward zero and the peak width $\Gamma$, which is resolution-limited below 7 K, becomes broader, indicative of the shorter lifetime of the excitations, as shown in the inset of Fig.~\ref{fig3}(c).  Figures~\ref{fig4}(a) and~\ref{fig4}(b) show the spin-wave dispersions obtained from several constant-energy and constant-${\bm Q}$ scans throughout the Brillouin zone along the two high-symmetry directions. The data points were obtained from resolution-convolution fits. Unfortunately, we were not able to determine the excitation energies of the high-energy modes owing to the high phonon background and low scattering intensity, which may be due to magnon instability.\cite{Chernyshev}

We analyze the low-energy spin-wave dispersion observed in Cs$_2$Cu$_3$SnF$_{12}$ in the framework of LSWT. The underlying spin structure used to calculate the spin-wave dispersion is that of the ${\bm q}\,{=}\,0$ structure for the uniform kagome lattice, in which all spins are oriented either toward or away from the center of a triangle (see Fig.\ref{fig1}(b)). In our previous study on Rb$_2$Cu$_3$SnF$_{12}$ (ref.~\onlinecite{Matan1}), we found that the DM interactions play a dominant role in singlet-triplet excitations, \textit{i.e.}, a large out-of-plane component of the DM vectors gives rise to large splitting between the $S^z\,{=}\,{\pm}\,1$ and 0 modes and reduces the energy gap at the $\Gamma$-point. Therefore, as a first approximation, we consider the DM interactions as the dominant anisotropy energy (referred to as the DM model), and express the spin Hamiltonian as

\small
\begin{equation}
{\cal H} = \sum_{\langle i,j\rangle} \left\{ J_{ij}\,({\bm S}_i \cdot {\bm S}_j) + {\bm D}_{ij}\cdot \left[{\bm S}_i \times {\bm S}_j \right]\right\}+J_2\sum_{\langle\langle k,l\rangle\rangle}({\bm S}_k \cdot {\bm S}_l),
\label{eq1}
\end{equation}
\normalsize
where $J_{ij}$ and $J_2$ are the nearest-neighbor (NN) and next-nearest-neighbor (NNN) exchange interactions, respectively, and ${\bm D}_{ij}$ are DM vectors. $J_{ij}$ are nonuniform as shown in Fig.~\ref{fig1}(b), and their magnitude is scaled by $J_{1}$, which can be written as $J_{1i}\,{=}\,a_iJ_{1}$ where $i\,{=}\,1, 2, 3$, and 4, while the strength of the DM vectors ${\bm D}_{ij}$ is scaled by the corresponding exchange interactions, $D^{\|}_{ij}\,{=}\,d_zJ_{ij}$ and $D^{\bot}_{ij} \,{=}\,d_{\rm p}J_{ij}$, where the configurations of the out-of-plane $(D^{\|}_{ij})$ and in-plane $(D^{\bot}_{ij})$ components of the DM vectors are illustrated in Fig.~\ref{fig1}(b). We neglect the interlayer interaction, because the triplet excitations in Rb$_2$Cu$_3$SnF$_{12}$ are dispersionless perpendicular to the kagome layer.\cite{Matan1} 

The LSWT calculations of the spin-wave dispersion as well as the scattering intensity for the DM model of eq.~(\ref{eq1}), which are shown in Fig.~\ref{fig4}(c) and Fig.~S2 in Supplemental Materials\cite{Supplement}, were performed using a symbolic algebra method written in \textit{Mathematica}. Details of the LSWT calculations are described in Supplemental Materials\cite{Supplement}. The results reveal 12 branches of spin-wave excitations, but only three dominant low-energy branches (Fig.~S2) are observed experimentally. The strong inelastic scattering intensity centered around ${\bm Q}_{\rm m}$ (Fig.~\ref{fig4}(c)) is consistent with the experimental data. The obtained fit parameters are $J_{1}\,{=}\,13.6(3)$ meV, $a_1\,{=}\,1$ (fixed), $a_2\,{=}\,1.0(1)$, $a_3\,{=}\,0.84(7)$, $a_4\,{=}\,0.70(5)$, $J_2\,{=}\,{-}\,1.07(2)$ meV, $d_z\,{=}\,{-}\,0.29(1)$ and $d_{\rm p}\,{=}\,0.057(4)$, giving $J^{\tiny \textrm{sw}}_{\tiny\textrm{avg}}\,{=}\,(J_{11}\,{+}\,J_{12}\,{+}\,J_{13}\,{+}\,J_{14})/4\,{=}\,12.1(7)$ meV. The solid lines in Figs.~\ref{fig4}(a) and~\ref{fig4}(b) represent the best fits with these parameters.  The splitting of the two higher energy modes ($\omega_1$ and $\omega_2$, see Figs.~\ref{fig4}(a) and~\ref{fig4}(b)) at the $\Gamma$-point results from zone folding due to the structural transition.  In the DM model, the energies of the $\omega_1$ and $\omega_2$ modes at the $\Gamma$-point are mainly determined by the out-of-plane component $D^{\|}$ of the DM vectors and the exchange interactions. The value of $D^{\|}$ is as large as $0.29J_{1i}$, which is the same order of magnitude as the value of $D^{\|}$ observed in Rb$_2$Cu$_3$SnF$_{12}$ (ref.~\onlinecite{Matan1}). This large out-of-plane component of the DM vectors stabilizes the ${\bm q}\,{=}\,0$ state, and thus is responsible for the magnetic ordering in Cs$_2$Cu$_3$SnF$_{12}$ as discussed by C\'epas \textit{et al.}\cite{Cepas} For a uniform kagome lattice, the in-plane component $D^\bot$ gives rise to the splitting of the $\omega_1$ and $\omega_2$ modes and the spin gap $\Delta$, which are expressed as $\omega_2\,{-}\,\omega_1\,{=}\,(2D^\bot D^{\|})/(J_1\,{+}\,J_2 )$ and $\Delta\,{=}\sqrt{3}D^\bot$, respectively. The large splitting of the $\omega_1$ and $\omega_2$ modes and the small spin gap $\Delta$ cannot be consistently described by the DM model with uniform $J_1$, attesting to the necessity of a spin model with the enlarged unit cell and nonuniform $J_{1i}$. The $\omega_1$ branch, which corresponds to the zero-energy mode in the absence of the DM interactions, is lifted considerably owing to the large $D^{\|}$. Its weak dispersion and lowest spin gap at the K-point can be ascribed to a small ferromagnetic next-nearest-neighbor interaction ($J_2\,{<}\,0$). Another possibility accounting for the dispersion of the $\omega_1$ mode is the quantum fluctuations, which are dominant for the spin-1/2 case and favor the $\sqrt{3}\,\,{\times}\sqrt{3}$ ordering at the K-point over the ${\bm q}\,{=}\,0$ ordering\cite{Sachdev,Matan1,Hwang2}.

Although the spin-wave dispersion observed in Cs$_2$Cu$_3$SnF$_{12}$ is qualitatively understandable in terms of LSWT and the DM model, there is a large quantitative disagreement between the exchange constant $J_{\tiny\textrm{avg}}$ obtained from the spin-wave dispersion ($J_{\tiny \textrm{avg}}^{\tiny \textrm{sw}}\,{=}\,12.1$ meV) and that obtained from the magnetic susceptibility data $J_{\tiny \textrm{avg}}^{\tiny\textrm{mag}}$. As shown by the solid line in Fig.~\ref{fig2}(a), the magnetic susceptibility is best described using $J_{\tiny \textrm{avg}}^{\tiny\textrm{mag}}\,{=}\,19.8$ meV when the interaction coefficients $a_i$, $J_2/J_1$ and $d_z$ are fixed, as those obtained from the spin-wave data with $d_{\rm p}\,{=}\,0$. Here, we neglected the small in-plane component of the DM vector $d_{\rm p}$. 
$J_{\tiny \textrm{avg}}^{\tiny\textrm{mag}}\,{=}\,19.8$ meV should be close to the true exchange constant. However, the dotted lines in Fig.~\ref{fig4}(a) and (b), which represent LSWT with $J_{\tiny\textrm{avg}}^{\tiny\textrm{mag}}$, show a large discrepancy between the LSWT result and the data especially for the $\omega_0$ mode.  We note that the slope of this mode is predominantly determined by $J_{\tiny \textrm{avg}}$. Therefore, we deduce that the quantum fluctuations decrease excitation energies from those obtained by LSWT, \textit{i.e.}, negative quantum renormalization of the excitation energies occurs in Cs$_2$Cu$_3$SnF$_{12}$. For a spin-1/2 triangular-lattice Heisenberg antiferromagnet, a recent theory predicts that at high energies spin-waves are strongly renormalized, so that the dispersion becomes flat\cite{Chernyshev,Starykh,Zheng}.  However, in contrast to the case of the triangular lattice, the renormalization factor $(R\,{=}\,J_{\tiny \textrm{avg}}^{\tiny \textrm{sw}} / J_{\tiny \textrm{avg}}^{\tiny\textrm{mag}}\,{=}\,0.61)$ in Cs$_{2}$Cu$_{3}$SnF$_{12}$ appears to be independent of the momentum transfer. 

We note the renormalization factors in other low-dimensional antiferromagnets. For Cu(DCOO)$_2$\,$\cdot$\,4H$_2$O, which is described as an $S\,{=}\,1/2$ square-lattice antiferromagnet, the positive quantum renormalization with $R\,{=}\,1.21$ was reported\cite{Ronnow}. This renormalization factor coincides with theoretical result\cite{Igarashi,Singh2}. For Cs$_2$CuCl$_4$, in which antiferromagnetic chains are coupled to form a spatially anisotropic triangular-lattice antiferromagnet, a large renormalization factor of $R\,{=}\,1.63$ was reported\cite{Coldea}. This large positive quantum renormalization is attributed not to the triangular geometry of the lattice but to the spinon excitations characteristic of antiferromagnetic chain\cite{Kohno}. For KFe$_3$(OH)$_6$(SO$_4$)$_2$, which is described as  an $S\,{=}\,5/2$ uniform KLAF, the renormalization factor is estimated as $R\,{=}\,0.90$ using the exchange constants determined from the dispersion relations\cite{Matan3} and magnetization and ESR measurements\cite{Fujita}. This fact together with the present result on Cs$_2$Cu$_3$SnF$_{12}$ shows that the negative quantum renormalization of the excitation energies is universal for KLAFs with an ordered ground state and enhanced with decreasing spin quantum number $S$.\\

This work was supported by a Grant-in-Aid for Scientific Research from the Japan Society for the Promotion of Science, and the Global COE Program funded by the Ministry of Education, Culture, Sports, Science and Technology of Japan. H.T. was supported by a grant from the Mitsubishi Foundation.  K.M. was supported by the Thailand Research Fund under grant no.\ MRG55800.\\[5mm]

\renewcommand{\thefigure}{S\arabic{figure}}
\setcounter{figure}{0}   
{\centering \textbf{Supplemental Materials to ``Large Negative Quantum Renormalization of Excitation energies in the Spin-1/2 Kagome Lattice Antiferromagnet"}}

\section*{Spin-wave calculations}
\begin{figure}[h]
\vspace{5mm}
\begin{center}
\includegraphics[width=8.5cm, clip]{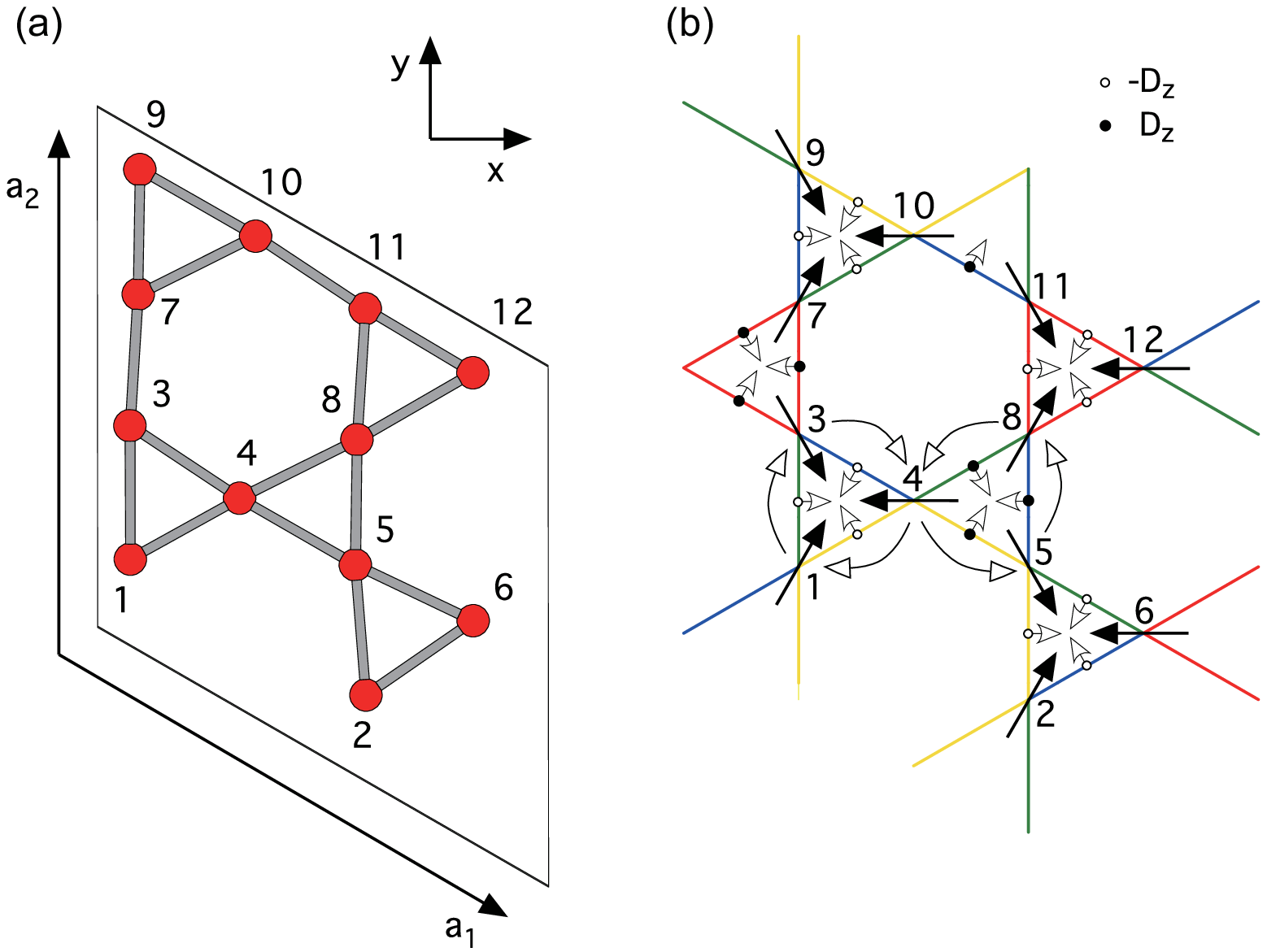}
\end{center}
\caption{(a) Twelve Cu$^{2+}$ $S\,{=}\,1/2$ spins form a distorted kagome plane. (b) Model of the ${\bm q}\,{=}\,0$ spin structure on a perfect kagome lattice used in the spin-wave calculations. Closed arrows denote spins while open arrows denote the DM vectors. The open and closed dots on the DM vectors denote the out-of-plane component; the open dots indicate the into-the-page direction and the closed dots indicate the out-of-page direction. Curved arrows denote the order of the cross product. }
\vspace{-6mm}
\label{figS1}
\end{figure}

At room temperature, Cs$_2$Cu$_3$SnF$_{12}$ crystallizes in the hexagonal structure (space group $R{\bar 3}m$) with the lattice parameters $a\,{=}\,7.142(4)$\,{\AA} and $c\,{=}\,20.381(14)$\,{\AA}\ \cite{Ono}. At $T_{\rm s}\,{=}\,185$ K, the system undergoes a structural transition. High-resolution time-of-flight powder neutron diffraction shows weak superlattice reflections, which indicate the enlarged $2a\,{\times}\,2a$ unit cell, and small splitting of the fundamental Bragg peaks, suggesting the change in crystal symmetry from hexagonal to monoclinic. However, as a good approximation, we retain the hexagonal system in the spin-wave analysis. In the enlarged unit cell, a two-dimensional unit cell comprises twelve Cu$^{2+}$ spins (Fig.\,\ref{figS1}(a)). Spin-wave excitations are calculated on the basis of the ${\bm q}\,{=}\,0$ spin structure on a perfect kagome lattice (Fig.\,\ref{figS1}(b)), where spins are oriented either toward the center of a triangle or away from it. The ${\bm q}\,{=}\,0$ structure is inferred from the magnetic Bragg reflections at $(2m, 2n, 0)$, where $m$ and $n$ are integers. We note that this spin structure is the same as that observed in the $S\,{=}\,5/2$ kagome lattice antiferromagnet KFe$_3$(OH)$_6$(SO$_4$)$_2$, and it is stabilized by the Dzyaloshinskii-Moriya (DM) interaction \cite{Nishiyama,Elhajal}.
As a first approximation, the spin Hamiltonian including the next-nearest-neighbor interaction and the DM interactions has the form
\begin{eqnarray}
{\cal H}\hspace{-1mm}&=&\hspace{-2mm}\sum_{\langle i,j\rangle, {\bm R}}\hspace{-1mm}\left\{J_{ij}\,({\bm S}_{i,{\bm R}}\,{\cdot}\,{\bm S}_{j,{\bm R}})+{\bm D}_{ij}\,{\cdot}\,[{\bm S}_{i,{\bm R}}\,{\times}\,{\bm S}_{j,{\bm R}}]\right\}\nonumber\\
&&+\ J_2\hspace{-2mm}\sum_{\langle\langle k,l\rangle\rangle, {\bm R}}\hspace{-2mm}({\bm S}_{k,{\bm R}}\cdot {\bm S}_{l,{\bm R}}), 
\end{eqnarray}
where the first sum is between nearest neighbors, the second sum is between second-nearest neighbors with the next-nearest-neighbor interaction $J_2$, and ${\bm R}$ is a lattice translational vector. The nearest-neighbor exchange interactions are shown in different colors in Fig.\,\ref{figS1}(b); green for $J_{11}$, yellow for $J_{12}$, red for $J_{13}$, and blue for $J_{14}$, where $J_{1i}\,{=}\,a_iJ_1$ ($i\,{=}\,1, 2, 3$, and 4). The curved arrows show the order of the cross product. If taken in the direction of the arrows, the cross product is positive and negative otherwise. The strength of the DM interaction for each bond is scaled by the corresponding nearest-neighbor exchange interaction, that is, $D^{\|}_{ij}\,{=}\,d_zJ_{ij}$ and $D^{\bot}_{ij}\,{=}\,d_{\rm p}J_{ij}$, where $D^{\|}_{ij}$ and $D^{\bot}_{ij}$ are the out-of-plane and in-plane components of the DM vectors, respectively.

The  lattice translation vectors of the two-dimensional unit cell are
\begin{eqnarray}
{\bm a}_1=a^{\prime}\left(\frac{\sqrt{3}}{2}, -\frac{1}{2}\right),\hspace{4mm} {\bm a}_2=a^{\prime}(0, 1)
\end{eqnarray}
where $a^{\prime}\,{=}\,2a$ is the in-plane lattice parameter of the enlarged unit cell. There are twelve spins in this enlarged unit cell located at
\begin{eqnarray}
 \left.
    \begin{array}{c}
{\bm d}_1\,{=}\,a^{\prime}(0, 0)\hspace{0.5mm},\hspace{3mm}{\bm d}_2\,{=}\,a^{\prime}\hspace{0mm}\left(\displaystyle\frac{\sqrt{3}}{4}, -\displaystyle\frac{1}{4}\right)\hspace{0mm},\vspace{3mm}\\
{\bm d}_3\,{=}\,a^{\prime}\hspace{0mm}\left(0, \displaystyle\frac{1}{4}\right)\hspace{0mm},
\hspace{3mm}{\bm d}_4\,{=}\,a^{\prime}\hspace{0mm}\left(\displaystyle\frac{\sqrt{3}}{8}, \displaystyle\frac{1}{8}\right)\hspace{0mm},\vspace{3mm}\\
{\bm d}_5\,{=}\,a^{\prime}\hspace{0mm}\left(\displaystyle\frac{\sqrt{3}}{4}\hspace{-1mm}, 0\right)\hspace{0mm},\hspace{3mm}{\bm d}_6\,{=}\,a^{\prime}\hspace{0mm}\left(\displaystyle\frac{3\sqrt{3}}{8}, -\displaystyle\frac{1}{8}\right)\hspace{0mm},\vspace{3mm}\\
{\bm d}_7\,{=}\,a^{\prime}\hspace{0mm}\left(0, \displaystyle\frac{1}{2}\right)\hspace{0mm},\vspace{1mm}\hspace{3mm}{\bm d}_8\,{=}\,a^{\prime}\hspace{0mm}\left(\displaystyle\frac{\sqrt{3}}{4}, \displaystyle\frac{1}{4}\right)\hspace{0mm},\vspace{3mm}\\
{\bm d}_9\,{=}\,a^{\prime}\hspace{0mm}\left(0, \displaystyle\frac{3}{4}\right)\hspace{0mm},\hspace{3mm}{\bm d}_{10}\,{=}\,a^{\prime}\hspace{0mm}\left(\displaystyle\frac{\sqrt{3}}{8}, \displaystyle\frac{5}{8}\right)\hspace{0mm},\vspace{3mm}\\
\hspace{0mm}{\bm d}_{11}\,{=}\,a^{\prime}\hspace{0mm}\left(\displaystyle\frac{\sqrt{3}}{4}, \displaystyle\frac{1}{2}\right)\hspace{0mm},\hspace{3mm}{\bm d}_{12}\,{=}\,a^{\prime}\hspace{0mm}\left(\displaystyle\frac{3\sqrt{3}}{8}, \displaystyle\frac{3}{8}\right)\hspace{0mm}.   
    \end{array}
  \right\}
\end{eqnarray}
$$\vspace{-2mm}\ $$

\begin{figure*}[tbh]
\begin{center}
\includegraphics[width=15.5cm, clip]{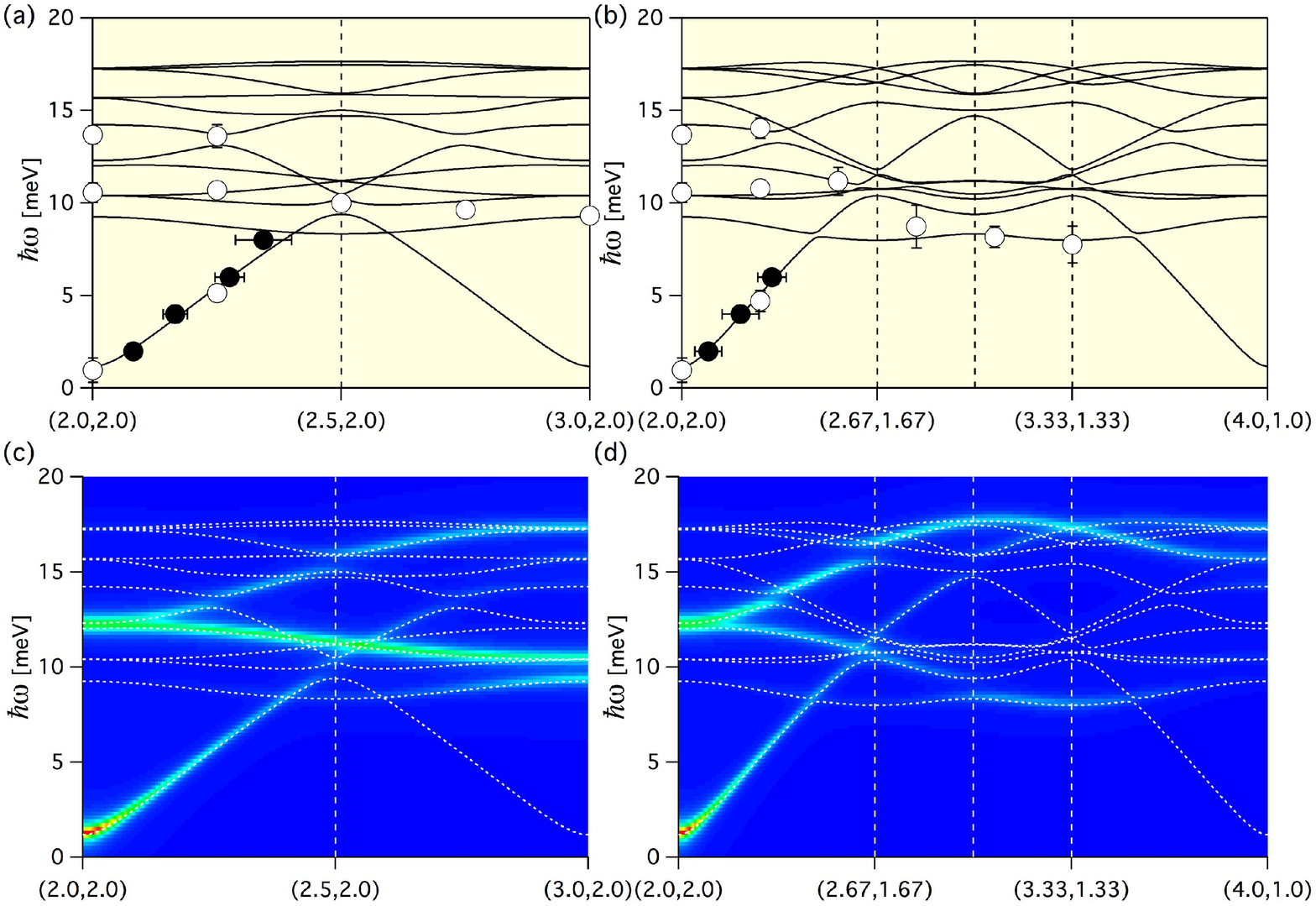}
\end{center}
\caption{Spin-wave dispersions along two high-symmetry directions. (a) and (b) show the spin-wave dispersions. Open and closed symbols denote the experimental data measured around $(2,2)$ and $(0,2)$, respectively. (c) and (d) show the scattering intensity. Red and blue indicate high and low intensities, respectively.}
\label{figS2}
\end{figure*}
We note that the lattice distortion is ignored and the perfect kagome lattice shown in Fig.\,\ref{figS1}(b) is considered. We express the spin interactions in terms of local axes, which are defined so that the $z$ axis coincides with the spin direction. The transformation matrices used for the projection of the spins in the kagome plane are defined as
\begin{eqnarray}
 \left.
    \begin{array}{c}
{\cal R}_1={\cal R}_2={\cal R}_7={\cal R}_8=\left[ \begin{array}{ccc}
0&\displaystyle\frac{\sqrt{3}}{2}&\displaystyle\frac{1}{2}\vspace{2mm}\\
0&-\displaystyle\frac{1}{2}&\displaystyle\frac{\sqrt{3}}{2}\vspace{2mm}\\
1&0&0
\end{array}
\right],\vspace{5mm}\\
{\cal R}_3={\cal R}_5={\cal R}_9={\cal R}_{11}=\left[ \begin{array}{ccc}
0&-\displaystyle\frac{\sqrt{3}}{2}&\displaystyle\frac{1}{2}\vspace{2mm}\\
0&-\displaystyle\frac{1}{2}&-\displaystyle\frac{\sqrt{3}}{2}\vspace{2mm}\\
1&0&0
\end{array}
\right],\vspace{5mm}\\
{\cal R}_4={\cal R}_6={\cal R}_{10}={\cal R}_{12}=\left[ \begin{array}{ccc}
0&0&-1\vspace{1mm}\\
0&1&0\vspace{1mm}\\
1&0&0
\end{array}
\right],
    \end{array}
  \right\}
\end{eqnarray}
where the rotation matrix ${\cal R}_i$ transforms the spin located at ${\bm d}_i$. However, the spins are canted owing to the DM interactions\,\cite{Yildirim}. The canting angle ${\eta}$ with respect to the kagome plane is given by
\begin{eqnarray}
{\eta}=\frac{1}{2}\cos^{-1}\left(1-\frac{2d_p^2}{3}\right). 
\end{eqnarray}
The rotation matrix ${\cal R}_c$ used to transform the local axes in the kagome plane to those with canting can be defined as
\begin{eqnarray}
{\cal R}_c=\left[\begin{array}{ccc}
\cos{\eta}\ &\ 0&\ -\sin{\eta}\vspace{1mm}\\
0\ &\ 1&\ 0\vspace{1mm}\\
\sin{\eta}\ &\ 0&\ \cos{\eta}
\end{array}\right] .
\end{eqnarray}
The combined rotation matrix ${\cal R}_i^{\prime}$ can be described by
\begin{eqnarray}
{\cal R}_i^{\prime}={\cal R}_i{\cal R}_c.
\end{eqnarray}
Hence, the transformation from spins in the local axes $\tilde{\bm S}_{i, {\bm R}}$ to those in the global axes ${\bm S}_{i, {\bm R}}$ is given by
\begin{eqnarray}
{\bm S}_{i, {\bm R}}={\cal R}_i^{\prime}{\tilde{\bm S}_{i, {\bm R}}}={\cal R}_i{\cal R}_c{\tilde{\bm S}_{i, {\bm R}}}.
\end{eqnarray}
The linearized Holstein-Primakoff transformations to the boson operators $c_i^{\dagger}$ (creation operator) and $c_i$ (annihilation operator) are given by
\begin{eqnarray}
\left.
\begin{array}{r}
{\tilde{S}_{i, {\bm R}}^x}=\frac{1}{2}[c_i^{\dagger}({\bm R})+c_i({\bm R})],\hspace{2mm} {\tilde{S}_{i, {\bm R}}^y}=\frac{{\rm i}}{2}[c_i^{\dagger}({\bm R})-c_i({\bm R})],\\
\ \vspace{-1mm}\\
{\tilde{S}_{i, {\bm R}}^z}=\frac{1}{2}-c_i^{\dagger}({\bm R})c_i({\bm R}).
\end{array}
\right.
\end{eqnarray}
The Fourier transforms of the boson operators are defined as
\begin{eqnarray}
\left.
\begin{array}{r}
c_i({\bm R})=\displaystyle\frac{1}{\sqrt{N}}\sum_{\bm k}c_i({\bm k}){\exp}\{{-{\rm i}{\bm k}\cdot({\bm R}+{\bm d}_i)}\},\\
\ \vspace{-1mm}\\
c_i^{\dagger}({\bm R})=\displaystyle\frac{1}{\sqrt{N}}\sum_{\bm k}c_i^{\dagger}({\bm k}){\exp}\{{{\rm i}{\bm k}\cdot({\bm R}+{\bm d}_i)}\}.
\end{array}
\right.
\end{eqnarray}

We have developed a symbolic algebra program in {\it Mathematica} to calculate the spin waves in Cs$_2$Cu$_3$SnF$_{12}$. The program performs the transformation to the local axes, the linearized Holstein-Primakoff transformations, and the Fourier transforms of the boson operators, and collects only the second-order terms, $c_i^{\dagger}({\bm k})c_i({\bm k})$, $c_i^{\dagger}({\bm k})c_i^{\dagger}(-{\bm k})$, $c_i(-{\bm k})c_i({\bm k})$, and $c_i(-{\bm k})c_i^{\dagger}(-{\bm k})$, to generate a $24\,{\times}\,24$ matrix, which is diagonalized to calculate the spin-wave dispersion.

Figure\,\ref{figS2} shows the spin-wave dispersions along two high-symmetry directions, which are described in Fig. 4(c), with the Hamiltonian parameters $J_1\,{=}\,13.6$ meV, $a_1\,{=}\,1.0,\ a_2\,{=}\,1.0,\ a_3\,{=}\,0.84,\ a_4\,{=}\,0.70,\ J_2\,{=}\,{-}\,1.07$ meV, $d_z\,{=}\,{-}\,0.29$, and $d_{\rm p}\,{=}\,0.057$. The resulting $J^{\tiny \textrm{sw}}_{\tiny\textrm{avg}}$ is equal to 12.1 meV. We found that $J^{\tiny \textrm{sw}}_{\tiny\textrm{avg}}$, which strongly depends on the slope of the spin-wave dispersion at small momentum transfer, is very robust while $J_1$, which is determined by the maximum excitation energy, can vary significantly. The large uncertainty of $J_1$ can be explained by the lack of experimental data at high energies.

The neutron scattering intensity $I({\bm Q}, {\omega})$ is calculated using the formula
\begin{eqnarray}
I({\bm Q}, {\omega})=I_0\sum_{{\alpha},{\beta}}\left({\delta}_{{\alpha}{\beta}}-{\hat Q_{\alpha}}{\hat Q_{\alpha}}\right)S^{{\alpha}{\beta}}({\bm Q}, {\omega}), 
\end{eqnarray}
where $I_0$ is a constant and $S^{{\alpha}{\beta}}({\bm Q}, {\omega})$ is the dynamic structure factor, which is given by
\begin{eqnarray}
S^{{\alpha}{\beta}}({\bm Q}, {\omega})={\delta}(\,{\hbar}{\omega}({\bm Q})\,{+}\,E_k\,{-}\,E_{k^{\prime}})\sum_{i,j}\langle S_i^{\alpha}({\bm Q})S_j^{\beta}(-{\bm Q})\rangle, 
\end{eqnarray}   
where ${\hbar}{\omega}({\bm Q})$ is the spin-wave energy. The correlation function $\langle S_i^{\alpha}({\bm Q})S^{\beta}_j(-{\bm Q})\rangle$ can be calculated from the eigenvectors of the $24\,{\times}\,24$ matrix obtained from the {\it Mathematica} program. The scattering intensity of the spin-wave excitations along the two high-symmetry directions (Figs.\,\ref{figS2}(c) and \ref{figS2}(d)) are calculated using Eqs. (11) and (12), where the $\delta$-function in Eq. 12 is replaced by a Lorentzian. The intensity is strong around $(2,2)$ and rapidly decreases toward $(3,2)$ and $(4,1)$, which is consistent with the experimental data (also see Fig. 4(c)). We note that the disagreement between the measured intensity of the ${\omega}_1$ and ${\omega}_2$ modes and the calculated scattering intensity (Figs.\,\ref{figS2}(c) and \ref{figS2}(d)) closed to the $\Gamma$-point could indicate the presence of other anisotropic terms such as the single-ion anisotropy\,\cite{Yildirim}, which can split the high-scattering-intensity modes.\vspace{2mm}\\

\bibliographystyle{apsrev}


\end{document}